\documentclass[twoside]{article}
\usepackage{fleqn,espcrc2,graphicx}
\def\simgt{\rlap{\lower 3.5pt\hbox{$\mathchar\sim$}}\raise 1pt\hbox {$>$}}
\def\simlt{\rlap{\lower 3.5pt\hbox{$\mathchar\sim$}}\raise 1pt\hbox {$<$}}
\title{$J/\psi$ and $\psi(2S)$ Production and Polarization at the 
 Tevatron\thanks{Talk presented at the {\sf 4th International Conference 
 on Hyperons, Charm and Beauty Hadrons\/, Valencia, Spain, 27-30 June 2000.
 To be published in the proceedings.}}}
\author{Michael Kr\"amer\\[3mm]
 Department of Physics and Astronomy\\
 The University of Edinburgh\\
 Edinburgh EH9 3JZ, Scotland}
\begin{document}
\begin{abstract}
The NRQCD factorization approach for quarkonium production at hadron
colliders is reviewed. The prediction of $J/\psi$ and $\psi(2S)$
transverse polarization at large transverse momentum is confronted
with recent experimental data, and potential shortcomings of the
theoretical analysis are discussed.
\vspace{1pc}
\end{abstract}
\maketitle
\section{Introduction}
The production of charmonium states at high-energy colliders has been
the subject of considerable interest during the past few years. New
results from $p\bar{p}$, $ep$ and $e^+e^-$ experiments have become
available, some of which revealed dramatic shortcomings of earlier
quarkonium production models. In theory, progress has been made on the
factorization between the short distance physics of heavy-quark
creation and the long-distance physics of bound state formation. The
colour-singlet model~\cite{Berger:1981ni} has been superseded by a
consistent and rigorous framework, based on the use of
non-relativistic QCD (NRQCD)~\cite{Bodwin:1995jh}, an effective field
theory that includes the so-called colour-octet mechanisms.  On the
other hand, the colour evaporation model~\cite{CEM1} of the early days
of quarkonium physics has been revived~\cite{CEM2}. However, despite
the recent theoretical and experimental developments the range of
applicability of the different approaches is still subject to debate,
as is the quantitative verification of factorization. Because the
charmonium mass is still not very large with respect to the QCD scale,
non-factorizable corrections~\cite{Brodsky:1997tv} may not be
suppressed enough and the expansions in NRQCD may not converge very
well.

In this situation cross checks between various processes, and
predictions of observables such as quarkonium polarization and
differential cross sections, are crucial in order to assess the
importance of different quarkonium production mechanisms, as well as
the limitations of a particular theoretical approach.  Among the
specific predictions of NRQCD, transverse polarization of direct
(i.e.\ not from $B$ and $\chi_c$ decays) $J/\psi$ and $\psi(2S)$
hadroproduction at large transverse momentum has emerged as the most
distinct and most accessible signature~\cite{Cho:1995ih}.  Transverse
polarization also discriminates NRQCD from other approaches, like the
colour evaporation model which predicts the quarkonium to be produced
unpolarized. In the following, I will briefly review the general
mechanism of quarkonium production at hadron colliders in the NRQCD
factorization approach. The prediction of $J/\psi$ and $\psi(2S)$
transverse polarization at large transverse momentum is confronted
with recent experimental data from the Tevatron, and potential
shortcomings of the theoretical analysis are discussed.

\section{Quarkonium production at hadron colliders} 
In the NRQCD approach, the cross section for producing a quarkonium
state $H$ at a hadron collider is written as a sum of factorizable
terms,
\begin{eqnarray}\label{eq_fac}
\mbox{d}\sigma(p\bar{p} \to H + X)&& \nonumber\\
&&\hspace*{-2.5cm} 
= \sum_n \mbox{d}\hat{\sigma}(p\bar{p} \to Q\overline{Q}\, [n] +
x)\,\langle {\cal{O}}^{H}\,[n]\rangle ,
\end{eqnarray} 
where $n$ denotes the colour, spin and angular momentum state of an
intermediate $Q\overline{Q}$ pair. The short-distance cross section
$\mbox{d}\hat{\sigma}$ can be calculated perturbatively in the strong
coupling $\alpha_s$. The NRQCD matrix elements $\langle {\cal O}^H[n]
\rangle$ (see \cite{Bodwin:1995jh} for their definition) are related
to the non-perturbative transition probabilities from the
$Q\overline{Q}$ state $n$ into the quarkonium $H$. They scale with a
definite power of the intrinsic heavy-quark velocity $v$
\cite{Lepage:1992tx}. ($v^2 \sim 0.3$ for charmonium and $v^2 \sim
0.1$ for bottomonium.) The general expression (\ref{eq_fac}) is thus a
double expansion in powers of $\alpha_s$ and $v$.

The NRQCD formalism implies that so-called colour-octet processes
associated with higher Fock state components of the quarkonium wave
function must contribute to the cross section. Heavy quark pairs that
are produced at short distances in a colour-octet state can evolve
into a physical quarkonium through radiation of soft gluons at late
times in the production process, when the quark pair has already
expanded to the quarkonium size.\footnote{Such a possibility is
ignored in the colour-singlet model, where only those heavy quark
pairs that are produced in the dominant Fock state (i.e.\ in a
colour-singlet state and with the spin and angular momentum quantum
numbers of the meson) are assumed to form a physical quarkonium.}

The production of $S$-wave charmonium in $p\bar{p}$ collisions at the
Tevatron has attracted considerable attention and has stimulated much
of the recent theoretical development in quarkonium physics.  The CDF
collaboration has measured cross sections for the production of
$J/\psi$ and $\psi(2S)$ states not coming from $B$ or radiative
$\chi_c$ decays, for a wide range of transverse momenta
$5\,\mbox{GeV}\; \simlt\; p_t(\psi) \;\simlt\;20\,\mbox{GeV}$
\cite{Abe:1992ww,Abe:1997jz}.  Surprisingly, the experimental cross
sections were found to be orders of magnitudes larger than the
theoretical expectation based on the leading-order colour-singlet
model~\cite{Baier:1983va}. This result is particularly striking
because the data extends out to large transverse momenta where the
theoretical analysis is rather clean and non-factorizable corrections
should be suppressed. The shortcoming of the colour-singlet model can be
understood by examining a typical Feynman diagram contributing to the
leading-order parton cross section, Fig.1(a). At large transverse
momentum, the two internal 
\begin{center}
\vspace*{-3.0cm}
\hspace*{-1.9cm}
\includegraphics[width=0.9\textwidth,clip]{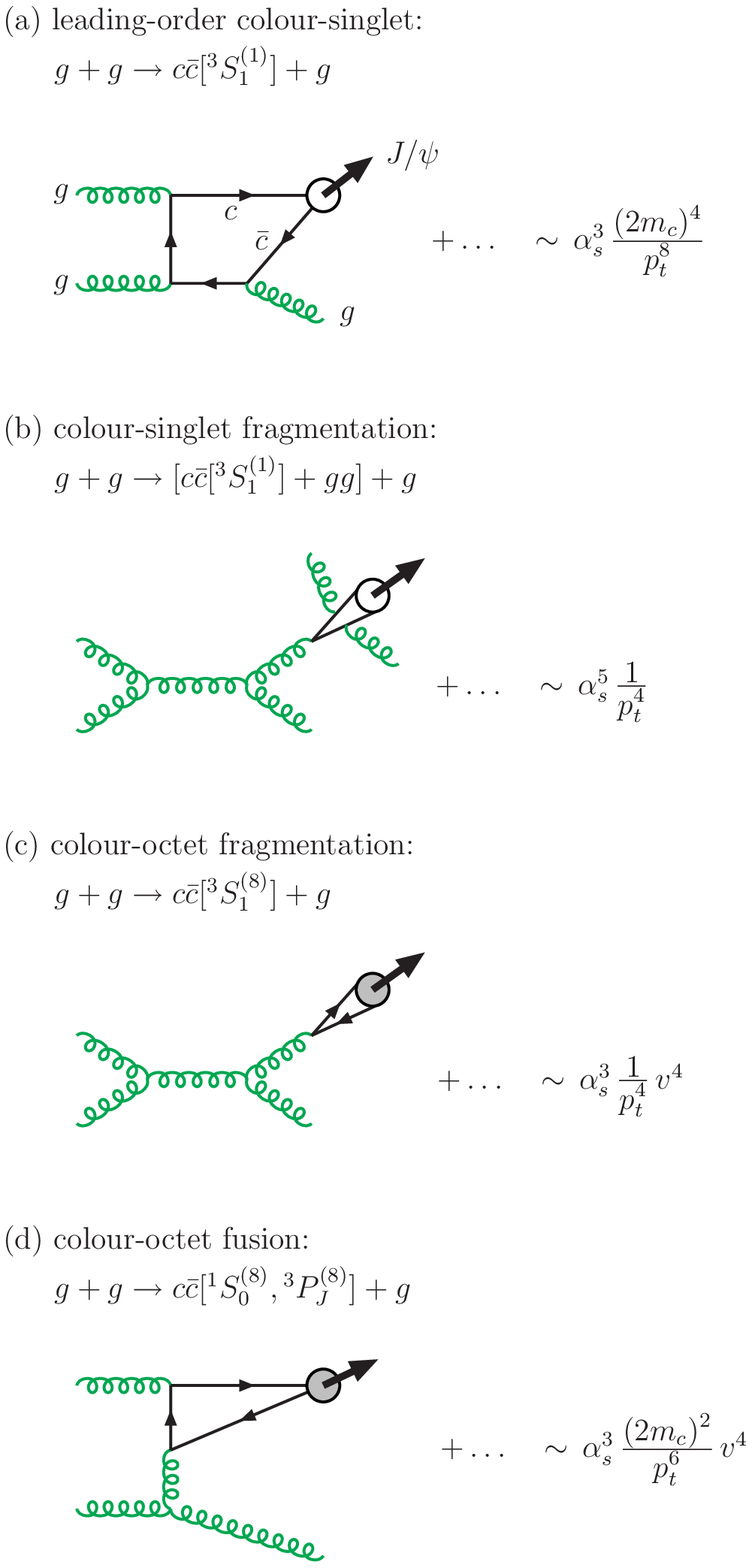}
\vspace*{-3.5cm}
\end{center}
\noindent
{\small Figure 1. Generic diagrams for $J/\psi$ production in hadron-hadron
collisions via colour-singlet and colour-octet channels.}\\[3mm]
quark propagators are off-shell by $\sim p_t^2$ so that the parton
differential cross section scales like $\mbox{d} \sigma/\mbox{d} p_t^2
\sim 1/p_t^8$, as indicated in the figure. 
On the other hand, when $p_t \gg 2m_c$ the quarkonium mass can be
considered small and the inclusive charmonium cross section is
expected to scale like any other single-particle inclusive cross
section $\sim 1/p_t^4$. The dominant production mechanism for
charmonium at sufficiently large $p_t$ must thus be via fragmentation
\cite{Braaten:1993rw}, the production of a parton with large $p_t$
which subsequently decays into charmonium and other partons.  A
typical fragmentation contribution to colour-singlet $J/\psi$
production is shown in Fig.1(b).  While the fragmentation
contributions are of higher order in $\alpha_s$ compared to the fusion
process Fig.1(a), they are enhanced by a power $p_t^4/(2m_c)^4$ at
large $p_t$ and can thus overtake the fusion contribution at $p_t \gg
2m_c$. When colour-singlet fragmentation is included, the $p_t$
dependence of the theoretical prediction is in agreement with the
Tevatron data but the normalization is still underestimated by about
an order of magnitude~\cite{Cacciari:1994dr}, indicating that an
additional fragmentation contribution is still missing. It is now
generally believed that gluon fragmentation into colour-octet
${}^3S_1$ charm quark pairs~\cite{Braaten:1995vv}, as shown in
Fig.1(c), is the dominant source of $J/\psi$ and $\psi(2S)$ at large
$p_t$ at the Tevatron. The probability of forming a $J/\psi(\psi(2S))$
particle from a pointlike $c\bar{c}$ pair in a colour-octet ${}^3S_1$
state is given by the NRQCD matrix element $\langle {\cal
O}^{J/\psi(\psi(2S))}[{}^3S_1^{(8)}] \rangle$ which is suppressed by
$v^4$ relative to the non-perturbative factor of the leading
colour-singlet term. However, this suppression is more than compensated
by the gain in two powers of $\alpha_s/\pi$ in the short-distance
cross section for producing colour-octet ${}^3S_1$ charm quark pairs
as compared to colour-singlet fragmentation. At ${\cal{O}}(v^4)$ in
the velocity expansion, two additional colour-octet channels have to
be included, Fig.1(d), which do not have a fragmentation
interpretation at order $\alpha_s^3$ but which become significant at
moderate $p_t\sim 2m_c$~\cite{Cho:1996vh}. The importance of the
${}^1S_0^{(8)}$ and ${}^3P_J^{(8)}$ contributions cannot be estimated
from naive power counting in $\alpha_s$ and $v$ alone, but rather
follows from the dominance of $t$-channel gluon exchange, forbidden in
the leading-order colour-singlet cross section.

The different contributions to the $J/\psi$ transverse momentum
distribution are compared to the CDF data \cite{Abe:1997jz} in
Fig.2. As mentioned above, the colour-singlet model at lowest order in
$\alpha_s$ fails dramatically when confronted with the experimental
results. When colour-singlet fragmentation is included, the prediction
increases by more than an order of magnitude at large $p_t$, but it
still falls below the data by a factor of $\sim 30$.  The CDF results
on charmonium production can be explained by including the leading
colour-octet contributions and adjusting the unknown non-perturbative
parameters to fit the data. Numerically one finds the non-perturbative
matrix elements to be of ${\cal{O}}(10^{-2}~\mbox{GeV}^3)$, see e.g.\
\cite{Nason:1999ta,Leibovich:2000qi}, perfectly consistent with the
$v^4$ suppression expected from the velocity scaling rules. Similar
conclusions can be drawn for $\psi(2S)$ production at the Tevatron.
\begin{center}
\vspace*{-1.2cm}
\hspace*{-0.65cm}
\includegraphics[width=0.55\textwidth,clip]{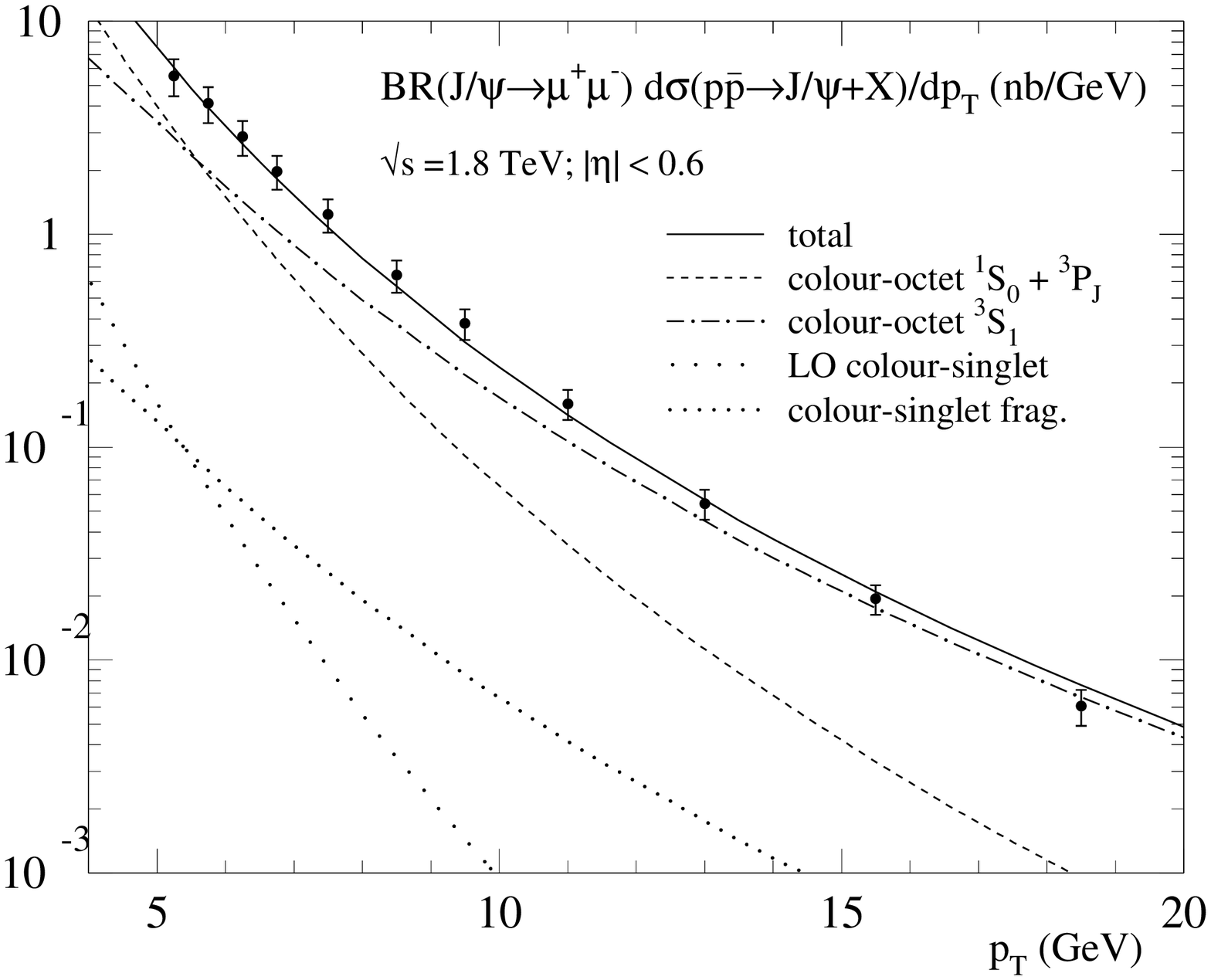}
\vspace*{-0.9cm}
\end{center}
\noindent
{\small Figure 2. Colour-singlet and colour-octet contributions to direct
$J/\psi$ production in $p\bar{p} \to J/\psi+X$ at the Tevatron
together with experimental data from CDF \cite{Abe:1997jz}. Parameters
and NRQCD matrix elements as specified in \cite{Nason:1999ta}.}\\

\section{$J/\psi$ and $\psi(2S)$ polarization at the Tevatron}
The analysis of the CDF data \cite{Abe:1992ww,Abe:1997jz} alone,
although very encouraging, does not strictly prove the
phenomenological relevance of colour-octet contributions because free
parameters have to be introduced to fit the data.  However, if
factorization holds the non-perturbative matrix elements extracted
from the $J/\psi$ and $\psi(2S)$ cross sections are universal and can
be used to make predictions for various processes and observables.
The single most crucial test of the NRQCD approach to charmonium
production at hadron colliders is the analysis of $\psi(2S)$
polarization at large transverse momentum. Recall that at large $p_t$,
$\psi(2S)$ production should be dominated by gluon fragmentation into
a colour-octet ${}^3S_1$ charm quark pair, Fig.1(c).  When $p_t \gg
2m_c$ the fragmenting gluon is effectively on-shell and transverse.
The intermediate $c\bar{c}$ pair in the colour-octet ${}^3S_1$ state
inherits the gluon's transverse polarization and so does the
quarkonium, because the emission of soft gluons during hadronization
does not flip the heavy quark spin at leading order in the velocity
expansion.  Consequently, at large transverse momentum one should
observe transversely polarized $\psi(2S)$~\cite{Cho:1995ih}.

The polarization can be measured through the angular distribution in
the decay $\psi\to l^+l^-$, given by $ \mbox{d}\Gamma /
\mbox{d}\!\cos\theta \propto 1+\alpha\,\cos^2\theta$, where $\theta$
denotes the angle between the lepton three-momentum in the $\psi$ rest
frame and the $\psi$ three-momentum in the lab frame. Pure transverse
polarization implies $\alpha=1$. Corrections to this asymptotic limit
due to higher order fragmentation contributions have been estimated to
be small~\cite{Beneke:1996yb}.  The dominant source of depolarization
comes from the colour-octet fusion diagrams, Fig.1(d), which are
important at moderate $p_t$. Still, at ${\cal{O}}(v^4)$ in the
velocity expansion, the polar angle asymmetry $\alpha$ can be
unambiguously calculated within
NRQCD~\cite{Beneke:1997yw,Leibovich:1997pa} in terms of the
non-perturbative matrix elements that have been determined from the
unpolarized cross section. In Fig.3 we display the theoretical
prediction ~\cite{Beneke:1997yw} (shaded band) for $\alpha$ in
$\psi(2S)$ production at the Tevatron as function of the $\psi(2S)$
transverse momentum, taking into account the ${}^1S_0^{(8)}$ and
${}^3P_J^{(8)}$ fusion channels and higher-order corrections to the
fragmentation contributions. No transverse polarization is expected at
$p_t\sim 5\,$ GeV, but the angular distribution is predicted to change
drastically as $p_t$ increases. A first measurement from
CDF~\cite{Affolder:2000nn} does not support this prediction, but the
experimental errors are too large to draw definite conclusions.

The analysis of $J/\psi$ polarization is complicated by the fact that
the experimental data sample~\cite{Affolder:2000nn} includes $J/\psi$
that have not been produced directly but come from decays of $\chi_c$
and $\psi(2S)$ mesons. The contribution from the radiative decays of
the higher excited states decreases, but does not eliminate, the
transverse $J/\psi$ polarization at large $p_t$~\cite{Braaten:1999qk}.
Again, the NRQCD factorization prediction is not supported by the
experimental data, see Fig.4.
\begin{center}
\vspace*{-1.2cm}
\hspace*{-0.25cm}
\includegraphics[width=0.50\textwidth,clip]{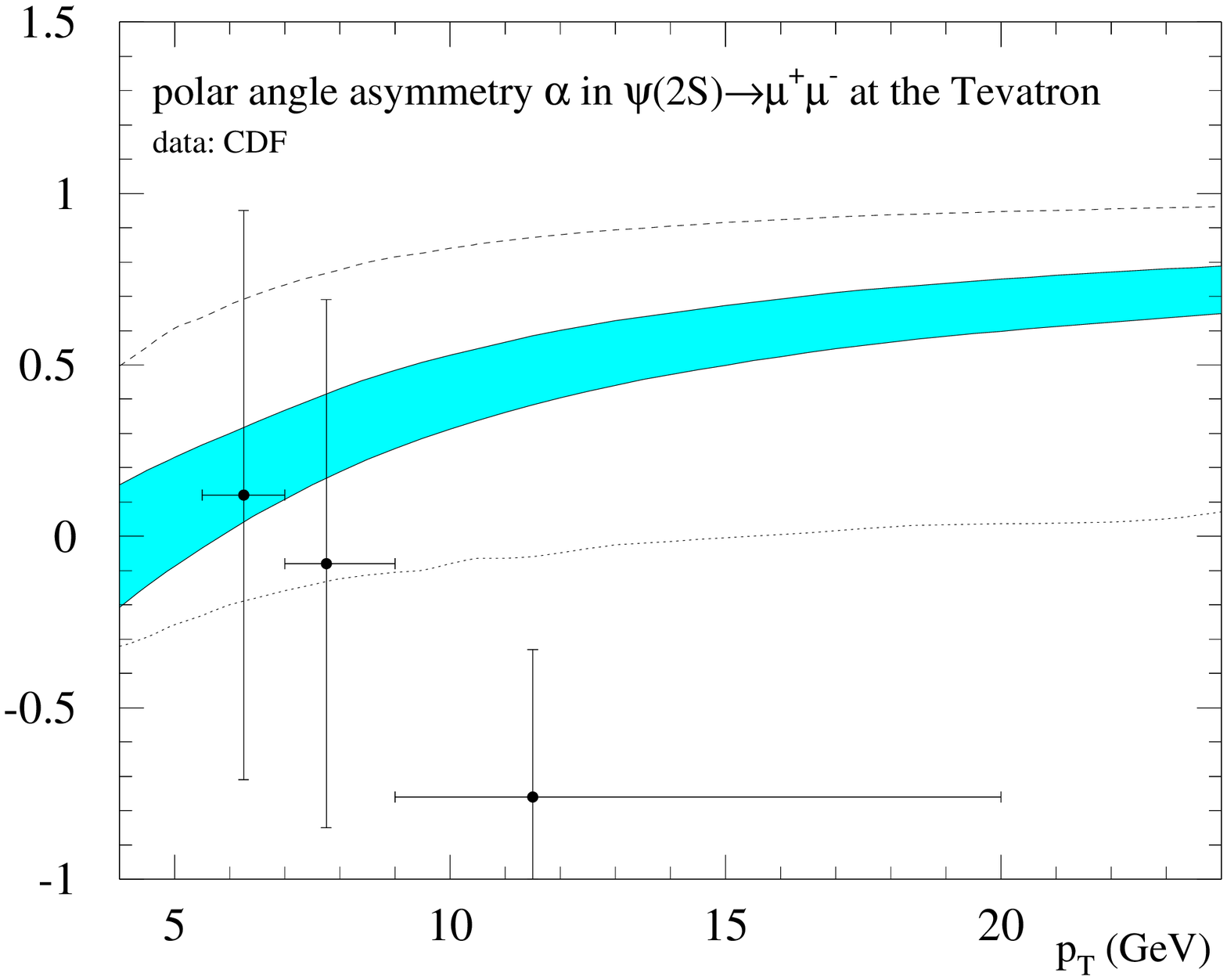}
\vspace*{-1.05cm}
\end{center}
\noindent
{\small Figure 3. Polar angle asymmetry $\alpha$ for direct $\psi(2S)$
production in $p\bar{p} \to \psi(2S)(\to \mu^+\mu^-)+X$ at the
Tevatron as a function of $p_t$ compared to data from CDF
\cite{Affolder:2000nn}. See~\cite{Beneke:1997yw} for details and
the choice of parameters. Also shown is the prediction for two extreme
choices of NRQCD matrix elements: $\langle {\cal
O}^{\psi}[{}^1S_0^{(8)},{}^3P_0^{(8)}] \rangle = 0$ (upper dashed
line) and and $\langle {\cal O}^{\psi}[{}^1S_0^{(8)}] \rangle = 0$,
$\langle {\cal O}^{\psi}[{}^3P_0^{(8)}] \rangle = 5\cdot
10^{-2}~\mbox{GeV}^3$, $\langle {\cal O}^{\psi}[{}^3S_1^{(8)}] \rangle
= 5\cdot 10^{-4}~\mbox{GeV}^3$ (lower dotted line), as discussed in
the text.}\\[-2mm]
\begin{center}
\hspace*{-0.1cm}
\includegraphics[width=0.44\textwidth,clip]{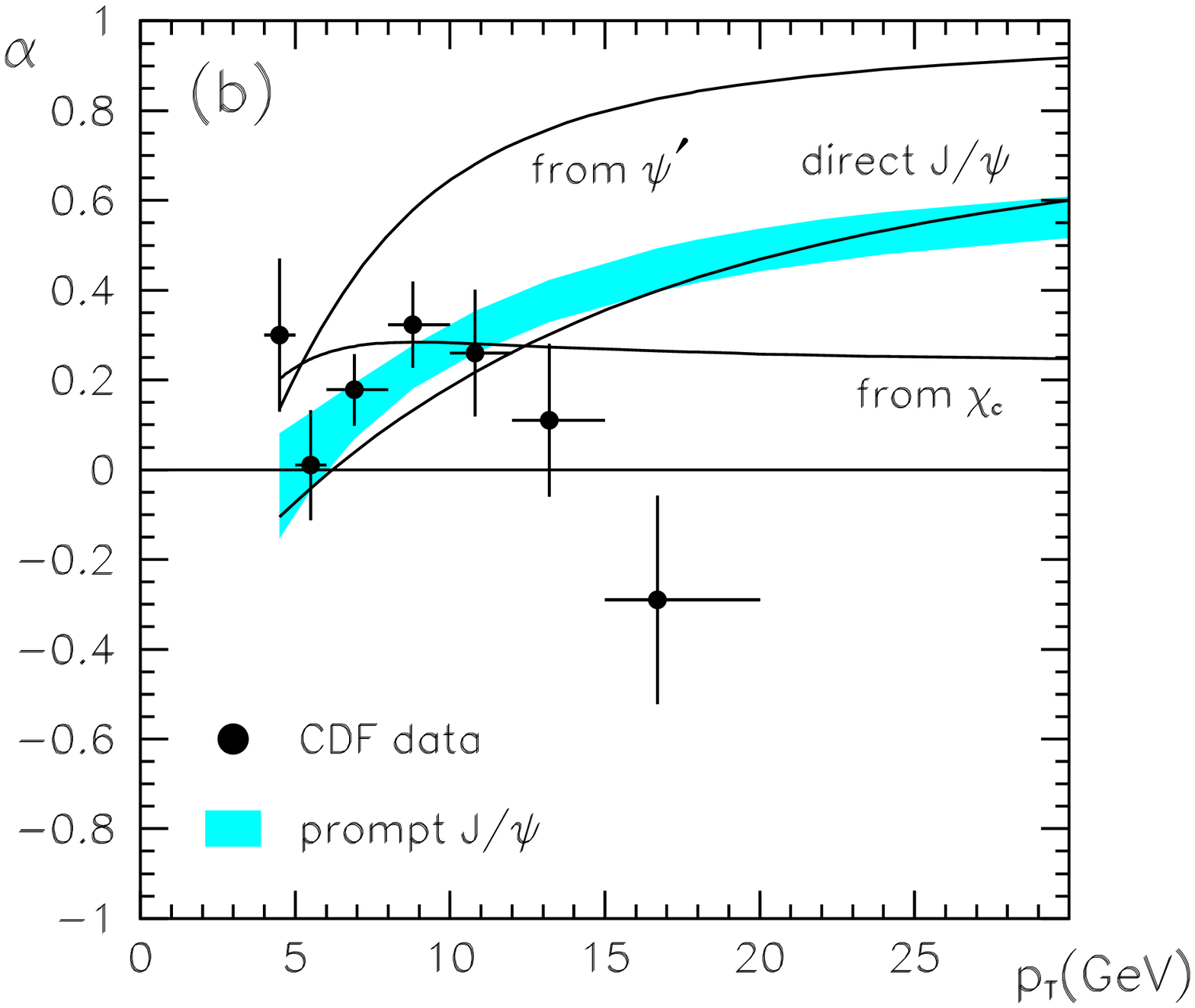}
\vspace*{-0.1cm}
\end{center}
\noindent
{\small Figure 4. Polar angle asymmetry $\alpha$ for prompt $J/\psi$
production (i.e.\ not from B-decay) in $p\bar{p} \to J/\psi(\to
\mu^+\mu^-)+X$ at the Tevatron as a function of $p_t$ compared to data
from CDF~\cite{Affolder:2000nn}. From \cite{Braaten:1999qk}.}
\newpage

\section{Discussion}
The absence of transversely polarized $J/\psi$ and $\psi(2S)$ at large
transverse momentum, if confirmed at the Tevatron Run~II, represents a
challenge for the application of NRQCD factorization to charmonium
production. In the following, I will discuss some possible
uncertainties and potential shortcomings of the present theoretical
analysis:\\[1mm]
$\bullet$ {\it Higher-order QCD effects} in the short-distance cross
section should be included to improve the accuracy of the theoretical
prediction, in particular in the intermediate $p_t$ region. While no
attempt has been made to estimate the impact of higher-order
corrections on the spin-dependence of the cross section, preliminary
studies indicate large NLO effects for the unpolarized
$p_t$-distributions~\cite{Petrelli:2000rh,Nason:1999ta}. These
corrections could strongly affect the determination of the NRQCD
matrix elements from the Tevatron data.  A full NLO analysis is
however needed before quantitative conclusions can be drawn. The
impact of non-vanishing transverse momentum of the incoming partons on
the determination of the NRQCD matrix elements has been analysed using
a Gaussian smearing of the $p_t$ distribution~\cite{Petrelli:2000rh},
Monte Carlo event generators~\cite{Cano-Coloma:1997rn} and, most
recently, the $k_t$-factorization
formalism~\cite{Yuan:2000cp,Hagler:2000eu}.  The actual size of the
effect, however, turns out to be very different for the different
approaches studied in the literature.\\
$\bullet$ The {\it uncertainty in the determination of NRQCD matrix
elements} translates into an uncertainty in the predicted yield of
transversely polarized $J/\psi$ and $\psi(2S)$. The extraction of
NRQCD matrix elements from the unpolarized cross section at the
Tevatron is affected by large theoretical uncertainties which have not
yet been fully quantified, see
e.g.~\cite{Nason:1999ta,Leibovich:2000qi}. In Fig.3 we also
show the polarization prediction for the extreme choices of vanishing
$\langle {\cal O}^{\psi}[{}^1S_0^{(8)},{}^3P_0^{(8)}]
\rangle$ matrix elements (upper dashed line) and a very small 
$\langle {\cal O}^{\psi}[{}^3S_1^{(8)}] \rangle$ matrix element (lower
dotted line, see Figure caption). The latter choice is motivated by
the recent analyses of $J/\psi$ and $\psi(2S)$ production at the
Tevatron in the $k_t$-factorization
approach~\cite{Yuan:2000cp,Hagler:2000eu} which predict a strongly
enhanced colour-singlet contribution and a substantially decreased
value of $\langle {\cal O}^{\psi}[{}^3S_1^{(8)}]
\rangle$~\cite{Hagler:2000eu}. A decreased value of the $\langle {\cal
O}^{\psi}[{}^3S_1^{(8)}] \rangle$ matrix element, relative to 
$\langle {\cal O}^{\psi}[{}^1S_0^{(8)},{}^3P_0^{(8)}]
\rangle$, would delay the onset of transverse polarization and reduce the
discrepancy between NRQCD predictions and data. However, such extreme
values for the NRQCD matrix elements would indicate a flaw in
our understanding of the velocity scaling rules~\cite{Lepage:1992tx}
which imply that all three matrix elements are of approximately equal
size ${\cal{O}}(v^4)
\approx {\cal{O}}(10^{-2}~\mbox{GeV}^3)$.\\
$\bullet$ {\it Heavy quark spin-symmetry} is violated by higher-order
terms in the NQRCD lagrangian and longitudinal polarization can arise
if the binding of the charm quark pair into $J/\psi$ and $\psi(2S)$
proceeds through two chromo-magnetic dipole transitions. According to
the velocity scaling rules these transitions are suppressed by $v^4$
and the corresponding corrections to transverse polarization should
not exceed $\approx 5\%$. Predictions of heavy quark spin-symmetry for
radiative charmonium decays agree quite well with experimental
data~\cite{Cho:1995ih} and there is no obvious reason why
spin-symmetry should fail when applied to charmonium
production. Still, the estimate of the spin-symmetry breaking
corrections to transverse polarization relies on the conventional
velocity scaling rules~\cite{Lepage:1992tx} which have not been firmly
established for the case of charmonium~\cite{Pineda:2000gz}.
Alternative velocity scaling rules~\cite{Beneke:1997av,Schuler:1997is}
imply a hierarchy of NRQCD matrix elements different from the standard
counting and might indicate larger spin-symmetry violating
corrections.\\
$\bullet$ Finally, charmonium production mechanisms {\it beyond NRQCD}
could be responsible for $J/\psi$ and $\psi(2S)$ depolarization. For
example, the leading-twist formalism of NRQCD factorization does not
include possible rescattering interactions between the intermediate
heavy quark pair and a comoving colour field. It has been shown that
such rescattering corrections could yield unpolarized
quarkonium~\cite{Marchal:2000wd}. The analysis~\cite{Marchal:2000wd},
however, relies on several simplifying assumptions, and further work
is needed to establish the importance of comover interactions for
charmonium production at large $p_t$.

\section{Conclusion}
The absence of a transverse polarization in $J/\psi$ and $\psi(2S)$
hadroproduction at large $p_t$ represents a serious challenge for the
application of the NRQCD factorization approach to charmonium
production. Several potential problems with the current theoretical
analysis have been discussed but no conclusive picture has emerged
yet.  If no transverse polarization is observed with higher statistics
at larger values of $p_t$ in Run~II of the Tevatron, one might have to
conclude that the charm quark mass is not large enough for a
non-relativistic approach to work in all circumstances.

\vspace*{4mm}

\small 

\noindent
{\it Note added}\\ After submission of this proceedings contribution,
charmonium polarization at the Tevatron has been addressed in the
$k_t$-factorization approach~\cite{Yuan:2000qe}.  The authors argue
that the ${}^1S_0^{(8)}$ production channel could be the dominant
source of $J/\psi$ and $\psi(2S)$ in the experimentally accessible
region of transverse momentum, which implies that quarkonium be
produced mainly unpolarized.

\vspace*{4mm}

\noindent
{\it Acknowledgements}\\ I would like to thank Martin Beneke for his
collaboration and the conference organizers for creating a pleasant
and stimulating atmosphere. A conference grant by the Royal Society is
gratefully acknowledged.

\end{document}